\documentclass[fleqn,10pt]{wlpeerj}

\usepackage{graphicx}
\usepackage{listings}
\usepackage{cite}
\usepackage{bchart}
\usepackage{enumitem}
\usepackage{booktabs}
\usepackage{multirow}
\usepackage{makecell}
\usepackage{listings}
\usepackage{lscape}

\usepackage{hyperref}

\title{Nanopublication-Based Semantic Publishing and Reviewing: A Field Study with Formalization Papers}

\author[1]{Cristina-Iulia Bucur}
\author[1]{Tobias Kuhn}
\author[2]{Davide Ceolin}
\author[1]{Jacco van Ossenbruggen}
\affil[1]{Vrije Universiteit Amsterdam, Amsterdam, The Netherlands}
\affil[2]{Centrum Wiskunde \& Informatica, Amsterdam, The Netherlands}
\corrauthor[1]{Cristina-Iulia Bucur}{c.i.bucur@vu.nl}

\begin{abstract}
With the rapidly increasing amount of scientific literature, it is getting continuously more difficult for researchers in different disciplines to keep up-to-date with the recent findings in their field of study. Processing scientific articles in an automated fashion has been proposed as a solution to this problem, but the accuracy of such processing remains very poor for extraction tasks beyond the most basic ones (like locating and identifying entities and simple classification based on predefined categories). Few approaches have tried to change how we publish scientific results in the first place, such as by making articles machine-interpretable by expressing them with formal semantics from the start. In the work presented here, we set out to demonstrate that we can formally publish high-level scientific claims in formal logic, and publish the results in a special issue of an existing journal. We use the concept and technology of nanopublications for this endeavor, and represent not just the submissions and final papers in this RDF-based format, but also the whole process in between, including reviews, responses, and decisions. We do this by performing a field study with what we call formalization papers, which contribute a novel formalization of a previously published claim. We received 15 submissions from 18 authors, who then went through the whole publication process leading to the publication of their contributions in the special issue. Our evaluation shows the technical and practical feasibility of our approach. The participating authors mostly showed high levels of interest and confidence, and mostly experienced the process as not very difficult, despite the technical nature of the current user interfaces. We believe that these results indicate that it is possible to publish scientific results from different fields with machine-interpretable semantics from the start, which in turn opens countless possibilities to radically improve in the future the effectiveness and efficiency of the scientific endeavor as a whole.
\end{abstract}

\begin{document}

\flushbottom
\maketitle
\thispagestyle{empty}

\section{Introduction} \label{introduction}
Considering the abundance of scientific articles that are published every day, keeping up with the latest research is becoming a significant challenge for researchers in many fields. This is at least partially due to the fact that we are still holding on to an archaic paradigm of scientific publishing: the canonical way to publish scientific results is by writing them up in long English texts called articles, which are in the best case easy to read by human experts but remain mostly inaccessible to automated approaches (except on a very superficial level with text mining approaches). These articles then undergo peer reviewing, which is typically done in a way that is secretive and not standardized, with the effect that the reviewing process may lack transparency and the valuable comments from the reviewers cannot be reused or build upon. There have been studies on the effectiveness of peer-reviewing in its current form \citep{Smith1988, Linkov2006, Kotturi2017} that showed not only systematic biases among peer-reviewers, but also a lack of transparency in the general peer-reviewing process as a whole \citep{Smith2010, Benda2011, Lee2012}. Making reviews open might alleviate some of these concerns by ensuring higher-quality reviews, while at the same time increasing the trust in the reviewing process and the quality of the scientific publications themselves.

A range of approaches have been proposed to address some of these problems by making scientific texts machine-readable, allowing for automatic summarising, finding and retrieving information easier and even the ability to (partially) reason on the scientific texts themselves. Text mining approaches work reasonably well when it comes to simple entity extraction with techniques like named-entity recognition to extract the main concepts from a text (e.g. \citep{AlMoslmi2020, Yadav2018}), but accuracy dramatically drops with more complicated tasks like relation extraction or identifying links between entities \citep{Etzioni2005, Xu2015, Zeng2014}.

The vast majority of existing approaches of making scientific texts machine-readable have one thing in common: they take the current paradigm of scientific articles for granted and therefore take them as their starting point to extract information. While it is important to try to process the vast amount of existing scientific literature that has the form of long English texts (and sometimes long texts in other languages), we should also think about how we can improve the way how we publish scientific insights in the first place. An important aspect of this is the vision of semantic publishing, which we mean here in the sense of \emph{genuine semantic publishing} \citep{Kuhn2017GenuineSP}, where the machine-interpretable formal semantics cover the main scientific claims the work is making. Nanopublications \citep{Groth2010}, which are small RDF-based semantic packages, have emerged as a powerful concept and technology for enabling such genuine semantic publishing.

In previous research we have applied nanopublications to implement a semantic and fine-grained model for reviewing \citep{Bucur2019ReviewModel}, and have extended this to semantically represent the full structure of (classical) scientific articles with their reviews and review responses as a single network of nanopublications \citep{Bucur2020}. In order to get closer to our vision of genuine semantic publishing, however, we need to represent not just the structure but also the main content of these articles, most importantly their main scientific claims. To that aim, we proposed in subsequent work the \emph{super-pattern}, a semantic template to represent the meaning of scientific claims in formal logic \citep{Bucur2021SP}.

Taking an example from our previous study as illustration of the super-pattern, it has been stated in the scientific literature \citep{Felix2002} that in particular kinds of cells in the rat brain (specifically, endothelial cells) some sort of stress called transient oxidative stress affects the expression of a protein called Pgp. The super-pattern consists of five slots that would in this example be filled in as follows:
\begin{itemize}[noitemsep]
\item Context class: rat brain endothelial cell
\item Subject class: transient oxidative stress
\item Qualifier: generally
\item Relation: affects
\item Object class: Pgp expression
\end{itemize}

Informally, we can read this in the following way: whenever there is an instance of transient oxidative stress in the context of an instance of a rat brain endothelial cell, then generally (meaning in at least 90\% of the cases), that instance of stress has the relation of affecting an instance of Pgp expression. Formally, it directly maps to this logic formula:
\begin{multline*}
P( \mbox{~} \exists z( \mbox{\sffamily\small pgp-expression}(z) \wedge \mbox{\sffamily\small in-context}(z,x) \wedge \mbox{\sffamily\small affects}(y,z) ) \mbox{~} | \\
\mbox{~} \mbox{\sffamily\small transient-oxidative-stress}(y) \wedge \mbox{\sffamily\small rat-brain-endothelial-cell}(x) \wedge \mbox{\sffamily\small in-context}(y,x) \mbox{~} ) \geq 0.9
\end{multline*}
This is stating in logic terms (in slightly non-standard notation using conditional probability as a shorthand) that given a thing $y$ of type \emph{transient-oxidative-stress} in the context of a thing $x$ of type \emph{rat-brain-endothelial-cell}, the probability of there being a $z$ of type \emph{pgp-expression} that is in the same context $x$ is at least 90\%. We have shown that this pattern can be applied to formalize most high-level claims found in scientific literature across disciplines \citep{Bucur2021SP}.

In the work to be presented below, we combine all these elements of our previous work --- namely semantic representation of reviews, scientific works as a networks of nanopublications, and representing the main claims with the super-pattern --- in order to implement genuine semantic publishing and putting it to the test in a field study. For practical reasons, we did not require the scientific claims in this field study to be novel ones, but they were selected from existing publications. This field study led to the publication of a special issue in an established journal (\href{https://www.iospress.com/catalog/journals/data-science}{Data Science}) at an established publisher (\href{https://www.iospress.com/}{IOS Press}). This special issue consists of what we call \emph{formalization papers}, which are nanopublication-based semantic publications whose novelty lies in the formalization of a previously published scientific claim.

In this research we therefore aim to answer the following research question:
\begin{itemize}
\item Are nanopublications and the super-pattern appropriate concepts to enable a new paradigm of scientific communication where authors publish their scientific findings with formal semantics?
\end{itemize}

The rest of this article is structured as follows. In Section \ref{background} we describe the current state of the art in the field of scientific publishing with regard to scientific knowledge representation, semantic publishing and semantic articles and also alternative proposed machine-readable approaches like nanopublications. In Section \ref{methodology} we describe our approach with regard to a new way of publishing, starting from a formal way of representing the content of scientific claims and ending with the representation of the publication process itself in what we call ``formalization papers''. We then report and discuss the results of the field study we performed using formalization papers in Section \ref{results}. Future work and conclusion of the present research are outlined in Section \ref{conclusion}.

\section{Background} \label{background}

We provide here the background on scientific knowledge representation, scientific publishing, and nanopublications in particular.

\subsection{Scientific knowledge representation} \label{knowledge-representation}

Novel proposals for the current ``Disruption Era'' \citep{Rahardja2019ScientificPM} include scientific publication management models that connect abstract knowledge with actual world problems in the constantly growing body of scientific knowledge \citep{Chi2018KnowledgeGI}, and the use of decentralized publication systems for open science using, for example, existing technologies like Blockchain and IPFS \citep{TenorioForns2019TowardsAD}.

With respect to the growth of scientific literature, there is a trend towards multidisciplinary and interdisciplinary research, together with an increased volume of publications every year \citep{Wong2019ACO}.
A range of methods have been proposed to make scientific articles more machine-readable: from structuring scientific works as Research Objects (RO) \citep{Bechhofer2013RO, Belhajjame2015RO} to using facets in order to uncover the main methods, data, code and other objects that are used in scientific articles \citep{Peroni2013Lenses, McGregor2008Facets}. Most approaches, however, have focused on automated content extraction from scientific articles as they are currently available. Recent machine learning techniques, for example, can after training with large sets of scientific articles extract the main concepts and structure of scientific articles \citep{Xu2015, Zeng2014}. While the results can be very valuable there are also clear limitations, with the resulting data needing almost always manual curation to achieve decent quality \citep{GarciaCastro2013, Coulet2011, Sernadela2015}.

A significant number of vocabularies and ontologies in many various domains have been developed, which are now ready to be used for scientific knowledge representation. However, they remain often difficult to find, access and understand due to the lack of documentation, versioning problems, and unresolvable URIs, among other things \citep{Garijo2020BestPF, Halpin2010, Hitzler2010, Jain2010}. A considerable amount of attention has also been given to the datasets accompanying scientific articles. The Data Set Knowledge Graph (DSKG), for example, covers datasets from over 600k scientific publications \citep{Frber2021TheDS}. An important development in this respect is the strong momentum behind the FAIR initiative to make research data Findable, Accessible, Interoperable, and Reusable \citep{Wilkinson2016TheFG}. A large amount of research is ongoing on how these FAIR principles can be put into practices (e.g. \citep{Garijo2020BestPF}).
Many other aspects of scientific communication have been approached with more formal representations, such as declaring authorship contributions with the Contributor Roles Taxonomy \citep{McNutt2018TransparencyIA} to mention just one of them.

Semantic technologies have been used extensively in the Life Sciences, e.g. for the representation and discovery of concepts, their relationships and associated supporting evidence in order to integrate distributed repositories \citep{Hannestad2021KnowledgeBW}. A variety of controlled vocabularies exist in these fields that can serve as the foundation to represent scientific knowledge in a structured way in order to semantically capture the context of scientific findings \citep{Chibucos2014, Slater2012, Madan2019}. The BEL language \citep{Slater2014} is one of the few attempts to represent the high-level scientific claims themselves, with coverage for specific kinds of biological relations. Also many other domains besides the Life Sciences have adopted the principles and technologies of Linked Data and the Semantic Web, for example to build interlinked, heterogeneous, and semantically rich datasets in Cultural Heritage \citep{Hyvnen2012PublishingAU} and to find, address, and sometimes even solve research problems in Digital Humanities in interactive ways \citep{Hyvnen2020UsingTS}.

\subsection{Semantic publishing and semantic papers} \label{semantic-papers}

Semantic publishing applies semantic technology to scientific publishing, and comes in many forms and does not always align with what we have introduced above as genuine semantic publishing \citep{Kuhn2017GenuineSP}.
Under this umbrella of semantic publishing, there are approaches that generate semantically-enriched data models from digital publications for the integration, sharing, management and data comparison between publications \citep{PerezArriaga2018AutomatedDO}, study the semantic annotation and enhancement of scholarly articles \citep{Shotton2009SemanticPT}, provide dynamic visualizations in semantically enhanced papers \citep{Senderov2016TheOB}, assess the versioning aspect of semantic publishing \citep{Papakonstantinou2018AssessingLD}, create a global-scale platform with a dataset metadata for automated ingestion, discover, and linkage \citep{Jacob2017DataworldAP}, and propose semantic and web-friendly HTML-based alternatives to the currently PDF-focussed scientific writing process \citep{Peroni2016ResearchAI}. Semantic enhancements of scientific articles can be used for semantic interlinking, interactive figures, re-orderable references and even summary creation \citep{Shotton2009AdventuresIS}, and workflows to convert regular scientific articles into Linked Open Data have also been investigated \citep{Sateli2016FromPT}. Other approaches like the compositional and iterative semantic enhancement (CISE) advocate for a process of automatic semantic enhancement with semantic annotations \citep{Peroni2017AutomatingSP}.
A key role in most of these approaches is played by the variety of existing ontologies covering many different aspects of scientific publishing, most importantly the Semantic Publishing and Referencing (SPAR) Ontologies \citep{Peroni2014TheSP}.

Further note-worthy approaches include the work to semantically represent the setup and results of scientific studies, which then allows for running meta-analyses in a semi-automated way, better research replication, and automated hypothesis generation \citep{Tiddi2020FosteringSM}, and the development of the Open Research Knowledge Graph \citep{Jaradeh2019}. The latter is an initiative that aims to make research articles machine-readable by expressing their main scientific entities as a semantically interconnected knowledge graph. This graph is populated by methods such as extracting scientific concepts from the abstracts of scientific articles with the help of annotators \citep{Brack2020}. 

\subsection{Nanopublications} \label{nanopubs}

Nanopublications \citep{Groth2010} are a specific concept and technology that deserves special attention here. They have been proposed to express scientific (and other kind of) knowledge in Linked Data as small independent publication packages. They allow for rich provenance and metadata and are structured as follows: the assertion part contains the main content of the nanopublication, such as a scientific claim, expressed as RDF triples. The provenance part of a nanopublication describes how the assertion came about, e.g. by linking to the scientific methods used to arrive at the finding. The publication information part, finally, contains metadata about the nanopublication as a whole, such as by whom and when it was created. Nanopublications can be used for scientific findings, but also for representing the other elements of the scientific workflow, such as reviews and method descriptions, and more generally any kind of small coherent set of RDF triples \citep{kuhn2013broadening}. It has been shown how nanopublications can be made reliable and immutable by identifying them with cryptographic Trusty URIs \citep{kuhn2015making}, and how this allows for a decentralized network of services and template-based user interfaces such as Nanobench \citep{kuhn2021semantic}.

\section{Methodology} \label{methodology}
In this section we describe the approach and methods we followed to investigate whether nanopublications and the super-pattern are suitable to achieve genuine semantic publishing.

\subsection{Approach} \label{approach}
In our approach, we committed to a number of features. First, we wanted the final contributions to be published as ``real'' papers in a real established journal. They should be fully semantically represented (in RDF) but also have classical views that makes them look like other papers. Like that, they should also seamlessly integrate with the existing bibliometric system and it should be straightforward to cite them in the classical way.

Second, we decided to fully focus on arguably the most interesting element of scientific articles, which happens to also be one of the most challenging to formally represent: the main scientific claims the article is making. Scientific articles have a large number of other interesting pieces of information, e.g. information about the used methods among many other things, but for the purpose of the study to be presented, we focus only on the main claims.

Third, in order to retain the flexibility and power of nanopublications, we decided to refrain from providing a custom-built and optimized user interface that hides the complexity and limits the flexibility. By using generic template-based nanopublication tools and by customizing them solely by providing the templates, we hoped to get a better understanding of how the nanopublication technology works for such kinds of content and workflows in general, and not just for our specific case. On the other hand, this also means that we were looking for a bit more technically minded authors who can handle interfaces that do not come with all the comfort of polished specific applications.

Fourth, we wanted to test a system that \emph{could} be used to publish novel claims, but decided for practical reasons to focus on formalizing claims from previously published articles. Our approach is therefore based on what we call \emph{formalization papers} that contribute novel formalizations of existing claims.

Finally, we wanted to cover not just these main claims, but the whole publishing workflow that involves the initial submission of contributions, their reviewing, the responses to the reviews, the updated versions, and the final decision, and represent these as independent but interlinked nanopublications.

\subsection{Formalization Papers}
\label{formalization-papers}

Our approach builds upon our new concept of formalization papers. A formalization paper contributes a semantic formalization of one of the main claims of an already published scientific article. Its novelty therefore lies solely in the formalization of a claim, not the claim itself. The authors of such formalization papers consequently take credit for the way how the formalization is done, but not for the original claim (unless that claim happens to come from the same authors).

\begin{figure}[tb]
	\centering
	\includegraphics[width=0.98\textwidth]{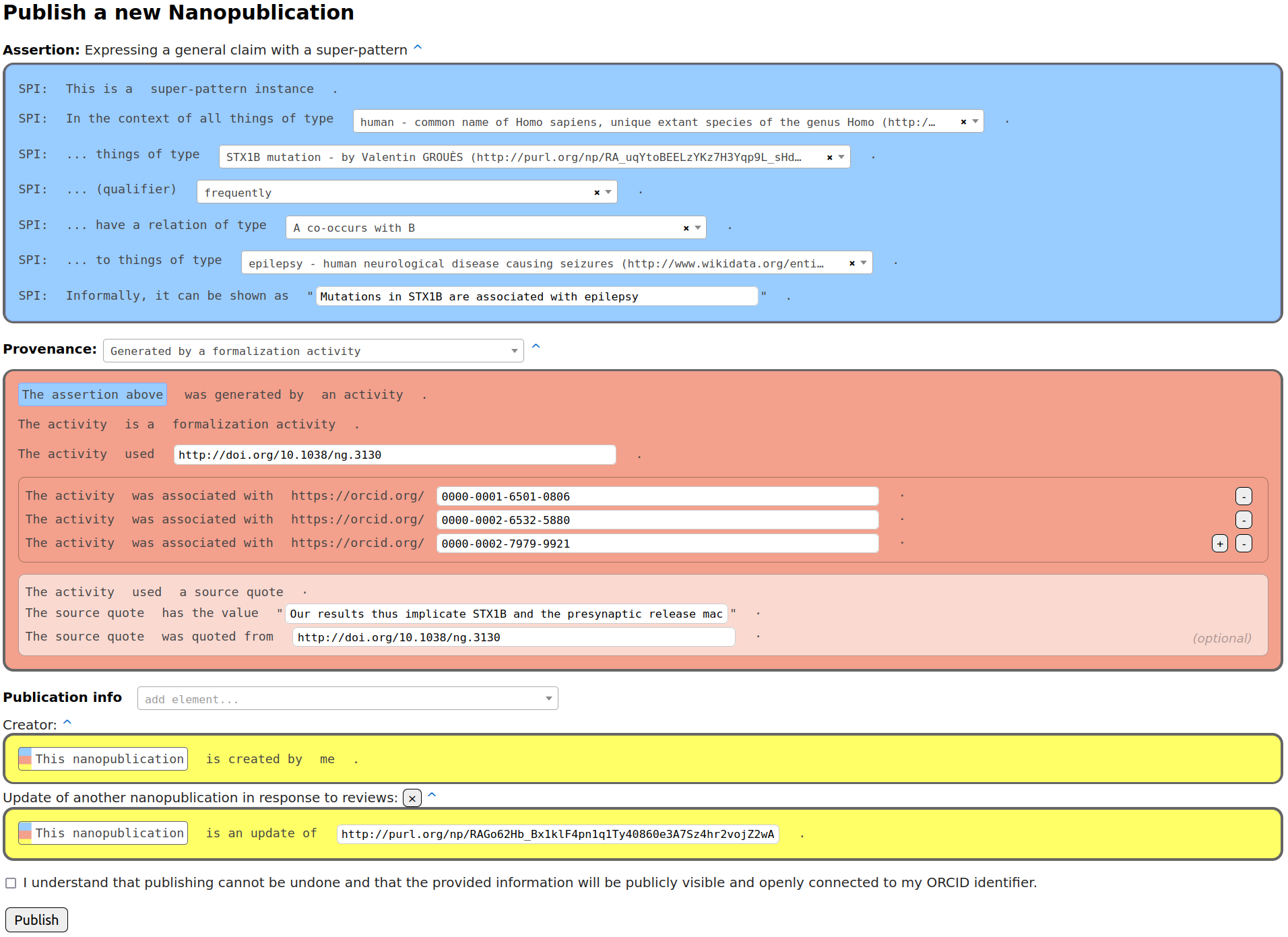}
	\caption{Formalization paper template from Nanobench as used by the participants of our study.}
	\label{fig:formalization-template}
\end{figure}

The content of a formalization paper is fully expressed in RDF in the form of nanopublications. Such a formalization paper can be shown in other formats to users, e.g. in HTML or PDF, but these are just views of the same underlying RDF content. Our formalization papers consist of nanopublications in which the assertion contains the formalization of the scientific claim using the super-pattern \citep{Bucur2021SP}, the provenance points to the original paper of the claim, and the publication information attributes the author of the formalization. Figure \ref{fig:formalization-template} shows an example of such a nanopublication in the interface the participants of our study used to create them. The instantiated super-pattern in the assertion part refers to a context class, a subject class, a qualifier, a relation type, and an object class according to the super-pattern ontology\footnote{\url{https://larahack.github.io/linkflows_superpattern/doc/sp/index-en.html}}. In the process of coming up with such a formalization, one often realizes that for some of the class slots of the super-pattern (i.e. context, subject, and object class) the class that should be filled in to arrive at a correct formalization is not directly defined in any existing vocabulary or ontology and as such, this class might need to be minted as well. The provenance part of the nanopublication describes the ``formalization activity'' that was conducted in order arrive at this formalization from what is written in the source publication. The precise phrase from that source publication that was used can be quoted too.

\subsection{Tools}

In order to publish formalization papers, class definitions, and all the other kinds of nanopublications (submissions, reviews, responses to reviews, and decisions), we use Nanobench \citep{kuhn2021semantic}\footnote{\url{https://github.com/peta-pico/nanobench}}. Figure \ref{fig:formalization-template} introduced above shows a screenshot of the publishing page of Nanobench. Publishing in Nanobench is based on templates, which are themselves expressed in nanopublications. The form shown in the screenshot is automatically generated based on the information found in several template nanopublications that we created and published for that purpose. All the application-specific behavior is therefore semantically represented in the templates, and Nanobench can flexibly be used for any other kind of data and workflow.

\begin{figure}[tb]
	\centering
	\includegraphics[width=0.98\textwidth]{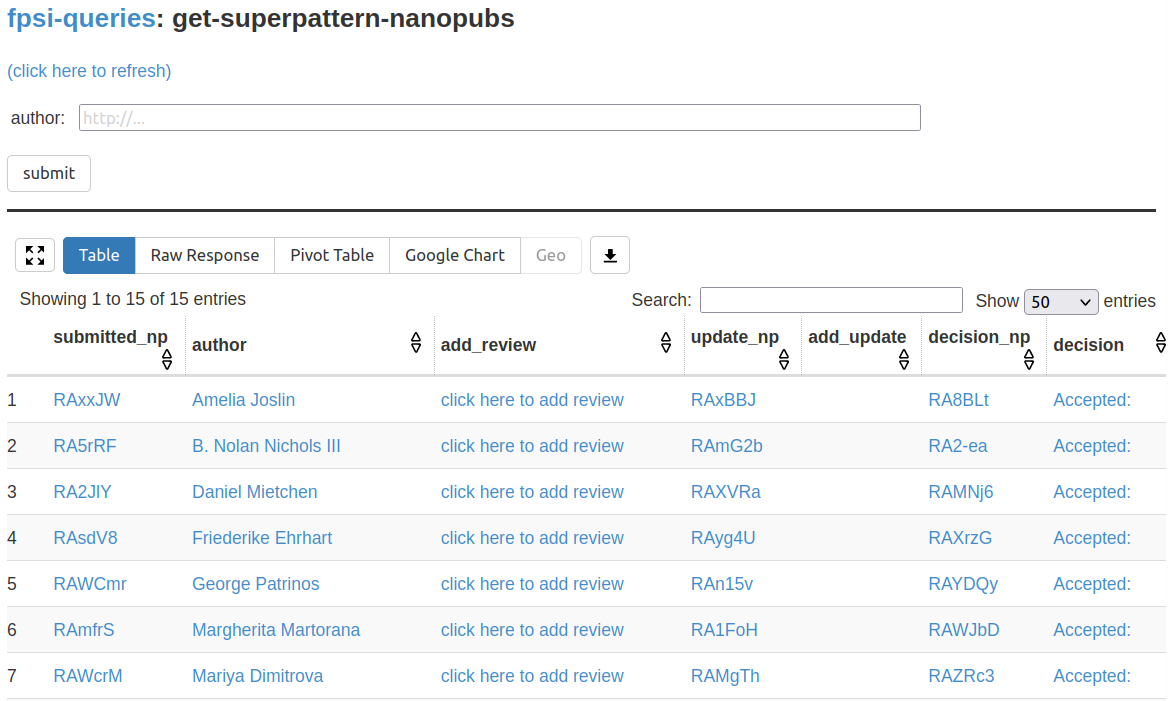}
	\caption{The Tapas interface listing submitted formalizations as the results of SPARQL queries over the nanopublication service network.}
	\label{fig:tapas}
\end{figure}

The second tool that we are using, Tapas \citep{lisena2019easy}\footnote{\url{https://github.com/peta-pico/tapas}}, is equally generic. It is a simple user interface component built on top of grlc \citep{merono2016grlc} that allows to run template-based SPARQL queries on RDF triple stores. In our case, we run it on SPARQL endpoints provided by the nanopublication service network \citep{kuhn2021semantic}. We use Tapas to show aggregations and overviews of submissions and reviews. Figure \ref{fig:tapas} shows a screenshot of the main submission overview. Tapas by itself is read-only, but we connect to the Nanobench tool with links that lead to partially filled-in forms (e.g. ``click here to add review'' in the screenshot).

\subsection{Field study design}
\label{field-study-design}

In order to test our approach, we devised a field study where interested authors could submit formalization papers, which upon acceptance are to be published as a special issue in the journal Data Science\footnote{\url{https://datasciencehub.net/}} by IOS Press. The goal of this was to demonstrate for the first time that scientific articles can be formalized and therefore machine-interpretable including the main scientific claims. As a secondary goal, we wanted to find out whether nanopublications are a good technology for that, and whether it is feasible to represent also the entire submission and reviewing process within the same framework.

Because the user interfaces we have at our disposal are still quite rough and technical, we restricted the set of possible authors and sent the call for papers on a by-invitation basis to selected groups of researchers who have previously worked or had experience with technologies like RDF and semantics.
We expect to be able to build more accessible user interfaces in the future that can show the inherent complexity in a way that does not require technical skills, but how this can be achieved is out of scope for this work.

Participants to our field study, thus the authors of formalization papers, formalized their own previously published claim, or a claim from a paper published by others. In the latter case, the formalization paper authors take credit for the formalization of the claim but not for the claim itself. All submissions to this special issue were peer-reviewed (also as nanopublications) using our previously proposed reviewing ontology \citep{Bucur2019ReviewModel}. Upon acceptance, these formalization papers are to be published in a journal at IOS Press, thereby giving them the same bibliometric status as other scientific articles, which leads to regular indexing in scientific article databases, counting of citations, and so on.

\begin{figure}[tbp]
  \noindent\makebox[\textwidth]{%
  \includegraphics[width=\textwidth]{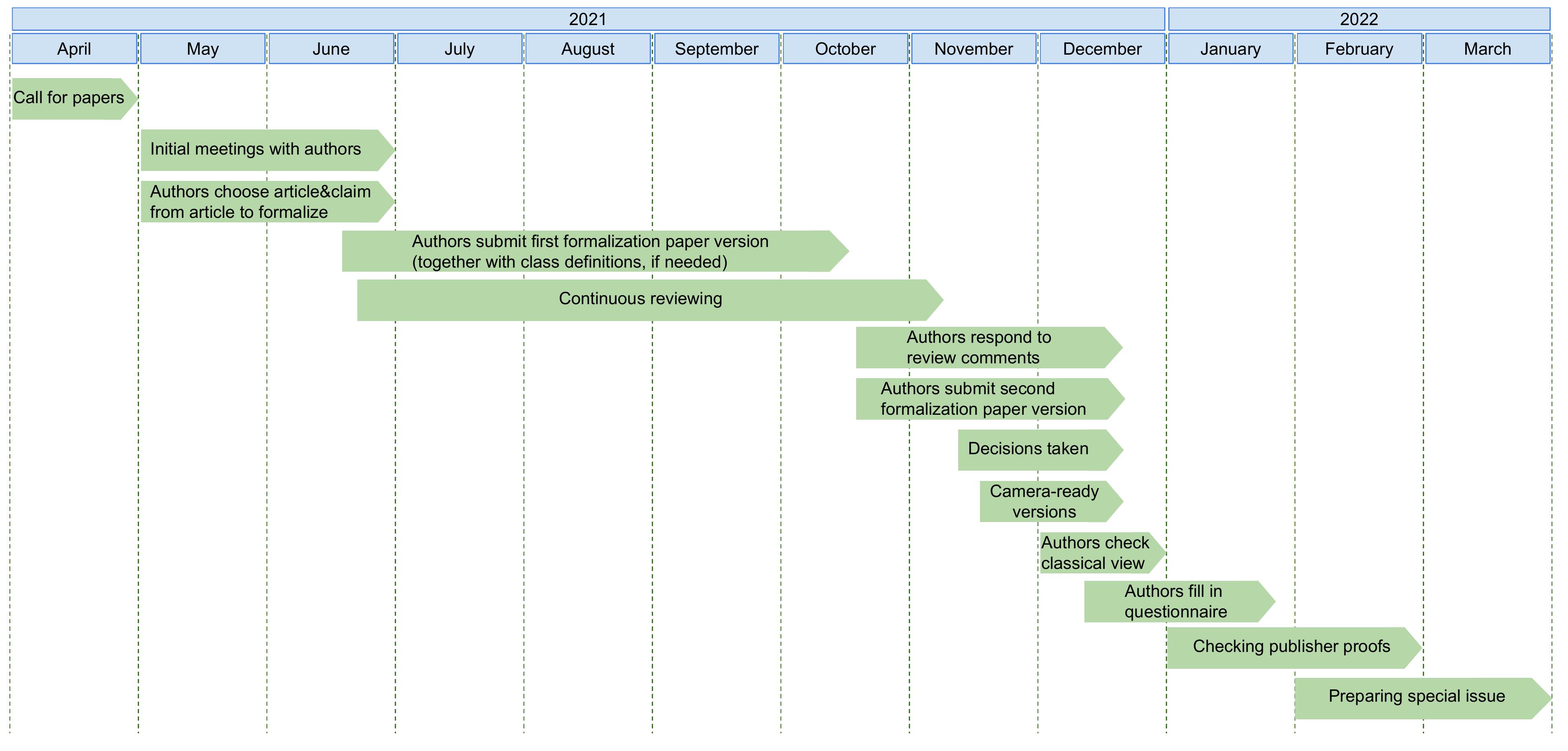}}
  	\caption{Timeline publication for the special issue with formalization papers at the Data Science journal.}
	\label{fig:special-issue-timeline}
\end{figure}

The whole timeline of the field study that encompasses the special issue can be seen in Figure \ref{fig:special-issue-timeline}.
The authors received close guidance on how to represent a claim of their choosing in RDF using the super-pattern and nanopublications, and on the various stages of the publication process. Authors took part in several information sessions and discussion meetings and were provided at each step with helper materials, videos, and even direct assistance if needed. In total, 24 such individual sessions were organized from May to December 2021.

In order to define a formalization, sometimes some of the class slots (i.e. context, subject, and object slots) of the super-pattern should be filled in with classes that are not yet defined in any existing vocabulary or ontology. In this case the authors first had to define these themselves, and they could do that also with the Nanobench tool loading a template for class definition. (Alternatively, they could also mint a new class identifier by other means, such as creating it on Wikidata.) The assertion of a nanopublication defining a new class may look for example as follows (\href{http://purl.org/np/RA_uqYtoBEELzYKz7H3Yqp9L_sHdU-kgL8R5EqmBsTVzE}{link to full nanopublication}):

{\footnotesize\begin{verbatim}
  sub:STX1B-mutation a owl:Class ;
    rdfs:subClassOf wd:Q42918 ;
    rdfs:label "STX1B mutation" ;
    skos:definition "mutation in STX1B" ;
    skos:relatedMatch wd:Q18048867 . 
\end{verbatim}}

Here, ``mutation'' from Wikidata (Q42918) is declared as super-class of the newly minted class ``STX1B mutation'', and ``STX1B'' (Q18048867) is linked as a related class.

Then the authors can publish their formalization in the form of a nanopublication using Nanobench (see Figure \ref{fig:formalization-template}), and afterwards they needed to submit it to the special issue using another Nanobench template, leading to an assertion like (\href{http://purl.org/np/RAWI_6Wpnnvn5scKXazYTqMftavW-HW9S-Alqlh1lf6Eo}{link to full nanopublication}):

{\footnotesize\begin{verbatim}
  <http://purl.org/np/RAGo62Hb_Bx1klF4pn1q1Ty40860e3A7Sz4hr2vojZ2wA>
    pso:withStatus pso:submitted ;
    frbr:partOf fpsi:DataScienceSpecialIssue . 
\end{verbatim}}

All submitted formalizations were subsequently reviewed. All authors were encouraged to review other submissions, and these reviews were semantic, open, and non-anonymous. These reviews were again done in nanopublications with the Nanobench tool. Such an example of a nanopublication assertion that contains a review modeled using the reviewing ontology can be seen below  (\href{http://purl.org/np/RAio--7IbPa3_ZSG3GspUsXeWP2ZwMIzy4Kzos0yZ7NIw}{link to full nanopublication}):

{\footnotesize\begin{verbatim}
  sub:comment a lfr:ReviewComment , lfr:ContentComment , lfr:NeutralComment ,
      lfr:SuggestionComment ;
    lfr:hasCommentText "Maybe the use of a causal relation like \"contributes to\"
                         can also be used here." ;
    lfr:hasImpact "1" ;
    lfr:refersTo <http://purl.org/np/RAGo62Hb_Bx1klF4pn1q1Ty40860e3A7Sz4hr2vojZ2wA> ;
    lfr:refersToMentioningOf sp:hasRelation .
\end{verbatim}}

In such a structured review (see more details in our previous research \citep{Bucur2020}), it is possible to specify various aspects that the review addresses including the aspect it comments on (syntax, style or content), the positivity/negativity of the review, the impact and the action that needs to be taken by the authors as the reviews see it (whether it is compulsory to be addressed, a suggestion or no action needs to be taken by the authors) and the importance of the point made by the review for the overall quality of the formalization. In the above example, the review comment makes a neutral point about the content of the given formalization with an importance of 1 out of 5, and is marked as a suggestion for the authors. The specific part of the formalization that this review targets is the \emph{sp:hasRelation} field, as indicated by the \emph{refersToMentioningOf} relation.

Subsequently, authors of the submissions could respond to the received review comments, again in nanopublications, and update their submissions based on these review comments. This is an example of a response to a review comment (\href{http://purl.org/np/RAAgR5ZKIIvujTwNwwxr6-bsjF1GXk_W7Zx7qxEeLrOX0}{link to full nanopublication}):

{\footnotesize\begin{verbatim}
  sub:comment a , lfr:ResponseComment lfr:DisagreementComment ,
      lfr:PointNotAddressedComment ;
    lfr:hasCommentText "I don't think the original publication shows a causal
                        relationship. It seems to me only a correlation is proven." ;
    lfr:isResponseTo
      <http://purl.org/np/RAio--7IbPa3_ZSG3GspUsXeWP2ZwMIzy4Kzos0yZ7NIw> ;
    lfr:refersTo <http://purl.org/np/RAeRSya2qIYymsBxiqOZP_oaQpHXUVXiydKvPCFM-7DDQ> .
\end{verbatim}}

This response registers the agreement with the point made by the reviewer (whether the author agrees totally, partially or not at all) and if that point was addressed, partially addressed or not addressed at all by the author. Moreover, a link to the respective review is given using the \emph{isResponseTo} relation, while the updated version of the formalization is indicated using the \emph{refersTo} relation. In our example, we see that the author does not agree with the point made by the reviewer and hence did not address the point raised by him, and also give a textual motivation on why this is the case.

\begin{figure}[tbp]
	\centering
	\includegraphics[trim=0 160mm 0 0, clip,width=\textwidth]{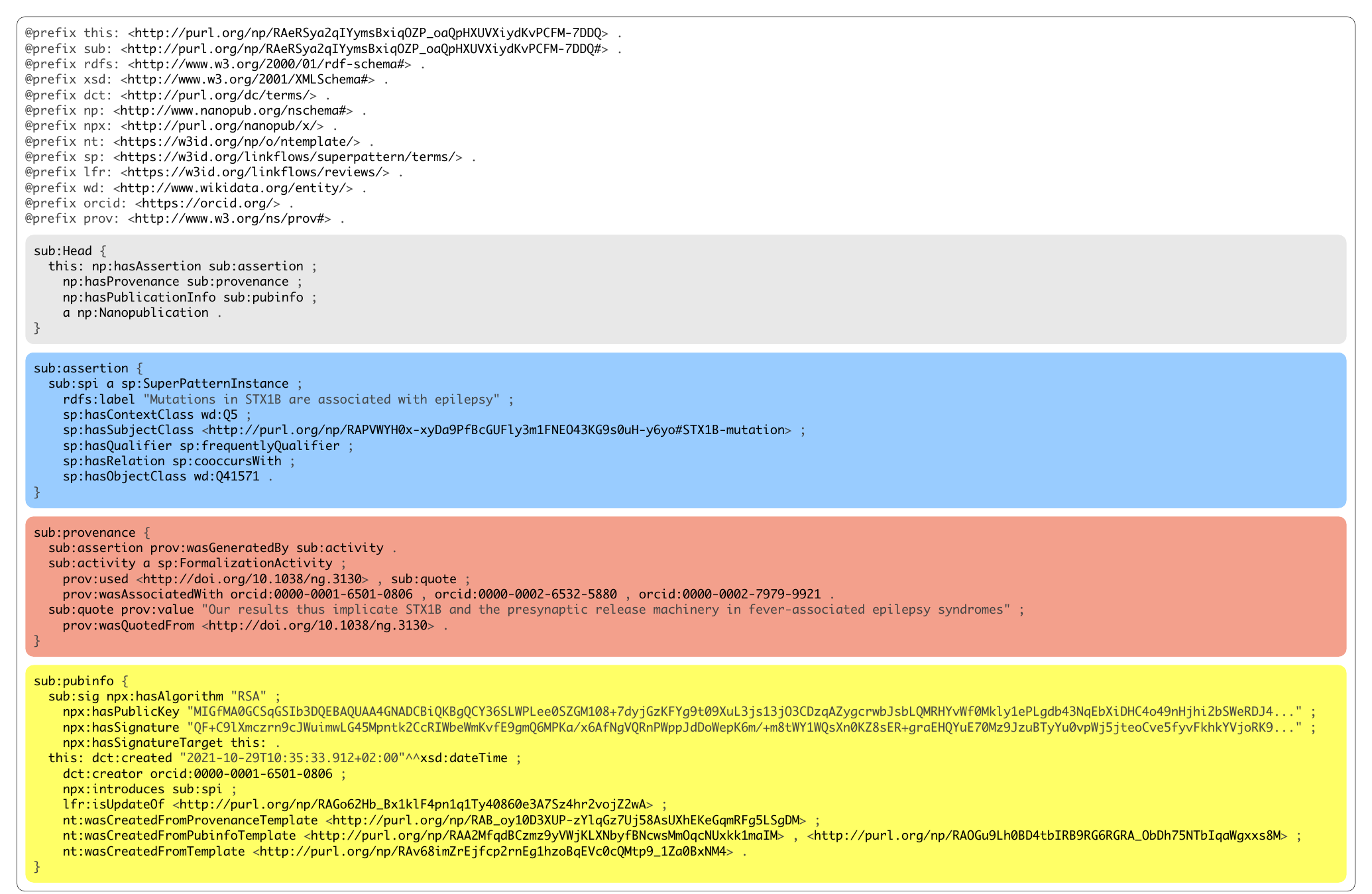}
	\caption{A \href{http://purl.org/np/RA22JAQihYeiJkNIjvwnxLPmjuG74yPcRXpPyVX8DV6fA}{nanopublication} containing a formalization paper.}
	\label{fig:FP-nanopub}
\end{figure}

Finally, the authors updated their formalizations with the same template as depicted in Figure \ref{fig:formalization-template}. The full final formalization nanopublication of the same example is shown in Figure \ref{fig:FP-nanopub}. For all updated submissions then a decision was made by us as the special issue editors about their acceptance. This decision was also represented as a nanopublication that looked as follows (\href{http://purl.org/np/RAKNnwB9sUaOdqUz3vk6FvIY8ckt5NsEn3scZb0MLux00}{link to full nanopublication}):

{\footnotesize\begin{verbatim}
  <http://purl.org/np/RAeRSya2qIYymsBxiqOZP_oaQpHXUVXiydKvPCFM-7DDQ>
    dct:description "All review comments were addressed and the formalization looks
                     good." ;
    pso:withStatus pso:accepted-for-publication ;
    frbr:partOf fpsi:DataScienceSpecialIssue . 
\end{verbatim}}

All formalizations reached a satisfactory level of quality, as indicated by the reviews and the authors' responses, and we therefore accepted all 15 submissions for publication.

\begin{figure}[tbp]
	\centering
	\includegraphics[width=0.6\textwidth]{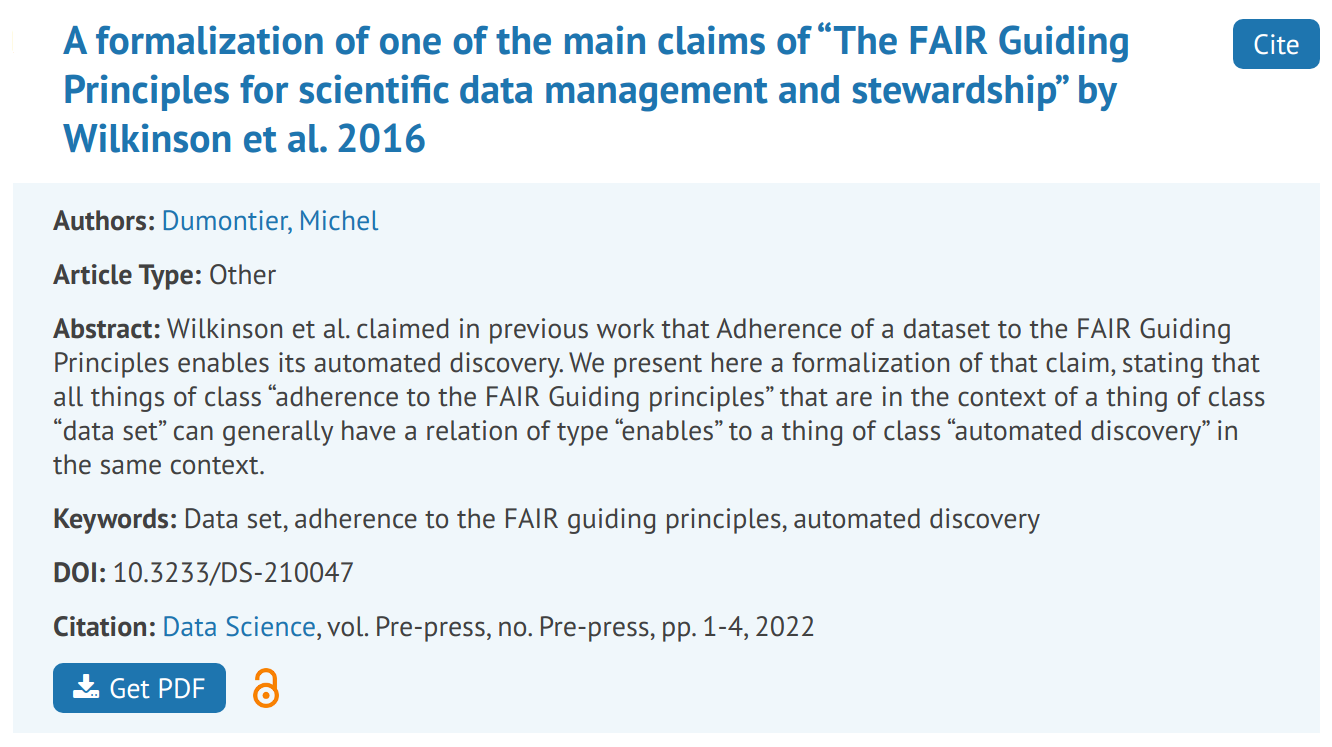}
	\caption{The ``classical view'' of a formalization paper, as it will appear on the publisher's website.}
	\label{fig:IOS-mock-up}
\end{figure}

In order to show the accepted papers in the special issue as if they were classical papers, to integrate them in the publisher's content management system, and to make them connect to the existing bibliometric system, we semi-automatically created ``classical views'' in the form of HTML and PDF versions of the nanopublications, as can be seen in Figure \ref{fig:IOS-mock-up}.

\subsection{User Feedback} \label{user-feedback}

In order to evaluate the general idea of formalization papers, all participants to the field study were asked to give us their opinion and report on their experiences about the involved processes and concepts. This evaluation was performed by means of a structured questionnaire consisting of four main parts, each one evaluating different aspects of the workflow.

In the first part, we are interested in assessing the difficulty of conceptually understanding the formalization paper idea and the super-pattern, and of performing the formalization tasks. In part two, we focus on the difficulty of the technical aspects in the various submission, reviewing and revision stages.
Part three addresses some more general aspects about the authors' experience and preferences. Authors were asked about their confidence in the formalization they published and about their interest of publishing such formalizations along their scientific publications in the future. We also asked them how important they think it is that all these steps are performed by the authors themselves (as they did). Moreover, in this part, authors could give us their opinion with regard to the importance of having a ``classical view'' along with the nanopublication representation of their formalization paper.
The fourth and final part of the questionnaire asked for the technical background of the authors. At the very end, the respondents could give further free-text feedback.
The full questionnaire is available in our supplemental material\footnote{\url{https://github.com/LaraHack/formalization_papers_supplemental/tree/main/questionnaire}}.

\section{Results} \label{results}

In this section we present the formalizations that resulted from our field study. We present a descriptive analysis of the generated data and analyze it also with the help of a network visualization. Finally, we report on the results from the user feedback questionnaire.

All the nanopublications that were created for all the submissions, formalization paper versions, the review comments, the responses to the review comments and the newly minted classes used in the formalizations together with the decisions are accessible online\footnote{\url{https://github.com/LaraHack/fpsi_analytics/tree/main/nanopubs}}, while the nanopublication index containing all these nanopublications has also been published\footnote{Nanopublication index: \url{http://purl.org/np/RAkLJW7vIsnKKJDf1iswdgtFPQSo3lEG_z8DhHfD7dofE}}. Also, the final submissions for the special issue with formalization papers at the Data Science journal \footnote{The link will be added in late March 2022, when the special issue is due for publication.} can be found online\footnote{\url{https://github.com/LaraHack/formalization_papers_supplemental/tree/main/accepted_submissions}}.

\subsection{Analysis of Formalizations}

In total, we had an initial number of 20 people that replied to our call for papers\footnote{\url{https://github.com/LaraHack/formalization_papers_supplemental/tree/main/call_for_papers}} from 12 different institutions from the United States of America, Germany, Luxembourg, Bulgaria, and The Netherlands from fields like biomedicine, bioinformatics, health sciences, ecology, data science, and computer science. After an initial information session, out of the 20 authors that responded to the call for papers, 18 decided to continue their participation. All these 18 authors that responded to the call for formalization papers managed in the end to publish (upon acceptance) their articles in a special issue at the Data Science journal.

We had a total of 15 formalization paper submissions, 13 with individual authors and 2 with joint authorship. Out of the total of 18 authors, two of these have both an individual submission and a joint-authorship one. The super-pattern instantiations of the final accepted formalization paper submissions can be seen in Table \ref{tab:submissions-superpattern}. Here, the classes used to instantiate the super-patterns that comprise the formalizations are given for each submission: the context, subject and object classes for each submission are listed, together with the qualifier and relations selected from the \href{https://larahack.github.io/linkflows_superpattern/doc/sp/index-en.html}{SuperPattern ontology} \citep{Bucur2021SP}. Each instantiation of the super-pattern can be interpreted as follows: ``Every thing of type [SUBJECT] that is in the context of a thing of type [CONTEXT] [QUALIFIER] has a relation of type [RELATION] to a thing of type [OBJECT] that is in the same context.''.

Looking at Table \ref{tab:submissions-superpattern}, we see that the super-pattern instances exhibit quite a broad variety of scientific fields (bioinformatics, biomedicine, pharmacology, data science, computer science) mostly linked to the life sciences. 7 out of the 15 submissions contain a formalization in which authors extracted a scientific claim from their own previously published article (submission number marked with $\diamond$). Additionally, out of the total 44 classes used in the formalizations, 22 new classes were minted using Nanobench (marked with *), while 4 were newly minted Wikidata classes (marked with **). 13 already-existing classes were reused from Wikidata (their Wikidata identifier is specified next to the class name) and 4 classes were referenced from other ontologies. 

\begin{table}[tbp]
    \centering
    \scriptsize
    \begin{tabular}{|p{0.5cm}|p{3.0cm}|p{3.0cm}|p{1.2cm}|p{1.1cm}|p{3.0cm}|}
        \hline
         & CONTEXT & SUBJECT & QUALIFIER & RELATION & OBJECT\\
         & (“in the context of all ...”) & (“things of type ...”) & & & (“to things of type...”)\\
        \hline
        \href{http://purl.org/np/RAxBBJ2WkonyQNlXfdCAOaCi64J_xqgVGeaLjVQow9M88}{1} & \href{http://purl.org/np/RAtsHwzNs36rGrLnoSbGrPD351Qw033Acoe4zmdXhsYlM#early-human-adipogenesis}{early human adipogenesis}* & \href{http://purl.org/np/RAxLYvJ1JrRf2JAowYGbGJleQPmqtpXnXsIvse7GmLeT8#regulatory-element-within-the-first-intron-of-FTO}{regulatory element within the first intron of FTO}* & \href{https://w3id.org/linkflows/superpattern/terms/generallyQualifier}{generally} & \href{https://w3id.org/linkflows/superpattern/terms/generallyQualifierhttps://w3id.org/linkflows/superpattern/terms/affects}{affects} & \href{http://purl.org/np/RAwkXiTv7qCtqOYzlR6ozZRGLRtG6mlogrYdRQ1E4dRDg#expression-of-genes-IRX3-and-IRX5}{expression of genes IRX3 and IRX5}* \\
        \hline
        \href{http://purl.org/np/RAmG2bXxwkIzARk4Mda-lqZU0RVnkpX7hUHBIPcdLHQUU}{2} & \href{http://www.wikidata.org/entity/Q101404862}{human motor neuron} (Q101404862) & \href{http://www.wikidata.org/entity/Q21133247}{TAR DNA binding protein} (Q21133247) & \href{https://w3id.org/linkflows/superpattern/terms/canGenerallyQualifier}{can generally} & \href{https://w3id.org/linkflows/superpattern/terms/contributesTo}{contributes to} & \href{http://purl.org/np/RAiUYY1dbEDbcsscapEmbMMHsgJmjEJ1yUoNsxZIH1r90#transcription-of-stmn2}{transcription of stmn2}* \\
        \hline
        \href{http://purl.org/np/RAXVRaFjWDlX5cZcVRXETaEIAx6QAyLK5JCrzDP-yDp9U}{3} $\diamond$ & \href{https://www.wikidata.org/wiki/Q107644116}{dejellied fertilizable stage VI Xenopus laevis oocyte}** & \href{https://www.wikidata.org/wiki/Q107644241}{strong static magnetic field}** & \href{https://w3id.org/linkflows/superpattern/terms/generallyQualifier}{generally} & \href{https://w3id.org/linkflows/superpattern/terms/affects}{affects} & \href{https://www.wikidata.org/wiki/Q5058180}{cell cortex (Q5058180)} \\
        \hline
        \href{http://purl.org/np/RAyg4UgIVovBGia-hk4qEuRzOq14fcOlYAclC6YGQaVYU}{4} $\diamond$ & \href{https://w3id.org/linkflows/superpattern/terms/UniversalContext}{(no context class)} & \href{https://www.wikidata.org/wiki/Q109406970}{genes associated with CAKUT}** & \href{https://w3id.org/linkflows/superpattern/terms/sometimesQualifier}{sometimes} & \href{https://w3id.org/linkflows/superpattern/terms/isSameAs}{is same as} & \href{https://www.wikidata.org/wiki/Q109406949}{targets of vitamin A}** \\
        \hline
        \href{http://purl.org/np/RAn15vsPJEVdJvjNKtBPo_oadtjeP9oc3Si-69FiJ4poQ}{5} $\diamond$ & \href{http://purl.org/np/RA9pwySo43TIfbvPuhK4ZuisvMsDvZ6TeR5N6MNKft8Nw#patient_undergoing_PCI}{patient undergoing PCI}* & \href{http://purl.org/np/RAOxICL4ULhzr5mxC9cyzStCBtpoETQGin6Vr-Ns7JNtA#pharmacogenomics_guided_clopidogrel_therapy}{pharmacogenomics guided clopidogrel therapy}* & \href{https://w3id.org/linkflows/superpattern/terms/generallyQualifier}{generally} & \href{https://w3id.org/linkflows/superpattern/terms/enables}{enables} & \href{http://purl.org/np/RAlfRfPak2jsyyVy4knjOmxQSYtociP8Cc0O7gemMtqQY#cost-effective_treatment}{cost-effective treatment}* \\
        \hline
        \href{http://purl.org/np/RA1FoHM9lwJ1XAV1eB871XcMAKfod73G_i4YtgoLpJVH0}{6} & \href{http://www.wikidata.org/entity/Q5}{human (Q5)} & \href{http://purl.obolibrary.org/obo/GO_0007224}{smoothened signaling pathway} & \href{https://w3id.org/linkflows/superpattern/terms/mostlyQualifier}{mostly} & \href{https://w3id.org/linkflows/superpattern/terms/affects}{affects} & \href{http://purl.obolibrary.org/obo/GO_0014002}{astrocyte development} \\
        \hline
        \href{http://purl.org/np/RABzhulhaPhOzo9MxWxl230N72-azdlpMNwu_HtDqsuUc}{7} $\diamond$ & \href{https://www.wikidata.org/wiki/Q28946370}{biodiversity data (Q28946370)} & \href{http://purl.org/np/RA5Txa3acYP9_MUWEw7s7wenDTB1QXNMB7UehJW-2E-_8#license-with-non-commercial-clause}{license with non-commercial clause}* & \href{https://w3id.org/linkflows/superpattern/terms/generallyQualifier}{generally} & \href{https://w3id.org/linkflows/superpattern/terms/inhibits}{inhibits} & \href{https://www.wikidata.org/wiki/Q58023280}{data reuse (Q58023280)} \\
        \hline
        \href{http://purl.org/np/RAMgThXW6xx8QiPmW9VhVuxWCN2ZWe-pmxDcFfdx_A7z0}{8} $\diamond$ & \href{http://purl.org/np/RAlm6vh2zpFLg189qrDYPtppkL790Pqaw-q2KUhyfJtRY#release-of-openbiodiv-knowledge-graph}{release of OpenBiodiv knowledge graph}* & \href{http://purl.org/np/RAaEkIiJLmBJP5kK3JdYjseCRqwutYbdnI8Q3VbzrK9VA#triple-in-openbiodiv-knowledge-graph}{triple in OpenBiodiv knowledge graph}* & \href{https://w3id.org/linkflows/superpattern/terms/generallyQualifier}{generally} & \href{https://w3id.org/linkflows/superpattern/terms/isSameAs}{is same as} & \href{http://purl.org/np/RAEpHUXRKtaLE3Z24sgIUdaxwTBsK2bjshyq9yF00145Y#semantic-triples-extracted-from-biodiversity-literature}{semantic triple extracted from biodiversity literature}* \\
        \hline
        \href{http://purl.org/np/RAXkuXJ4IK10Ai9F39_tOFDy6ewi7znau6OQhUEXP4nPc}{9} & \href{http://www.wikidata.org/entity/Q18036664}{UNC13A (Q18036664)} & \href{http://www.wikidata.org/entity/Q21133247}{TAR DNA binding protein (Q21133247)} & \href{https://w3id.org/linkflows/superpattern/terms/generallyQualifier}{generally} & \href{https://w3id.org/linkflows/superpattern/terms/inhibits}{inhibits} & \href{http://purl.obolibrary.org/obo/VariO_0504}{inclusion of cryptic exon} \\
        \hline
        \href{http://purl.org/np/RA22JAQihYeiJkNIjvwnxLPmjuG74yPcRXpPyVX8DV6fA}{10} $\diamond$ & \href{https://www.wikidata.org/wiki/Q1172284}{data set (Q1172284)} & \href{http://purl.org/np/RAodU4AmRjfzyjwtJK3luO0iyRJJPUBjkijKWdlMHvack#adherenceToTheFAIRGuidingPrinciples}{adherence to the FAIR guiding principles}* & \href{https://w3id.org/linkflows/superpattern/terms/canGenerallyQualifier}{can generally} & \href{https://w3id.org/linkflows/superpattern/terms/enables}{enables} & \href{http://purl.org/np/RAFQovt9yQD7nZ2tdZ9_Uhpb7CsfT3k64pK7dh63xd-50#automatedDiscovery}{automated discovery}* \\
        \hline
        \href{http://purl.org/np/RA12lVwEtmddK9OwDkZQZlgJaOD2-0NXtAtO_jDaG-3VQ}{11} & \href{http://www.wikidata.org/entity/Q5}{human (Q5)} & \href{http://purl.obolibrary.org/obo/MONDO_0014109}{NGLY1 deficiency} & \href{https://w3id.org/linkflows/superpattern/terms/alwaysQualifier}{always} & \href{https://w3id.org/linkflows/superpattern/terms/isCausedBy}{is caused by} & \href{http://purl.org/np/RAZVLqlkbwiX40n0GNxcxJany2Cw3oxMCrNuZtjBClryU#Dysfunction_of_ERAD_pathway}{dysfunction of ERAD pathway}* \\
        \hline
        \href{http://purl.org/np/RAbWbJCYlLhlYBDn9PVxdJP_WUbbi058aRcK-3sOJsRwY}{12} & \href{https://www.wikidata.org/wiki/Q874405}{social group (Q874405)} & \href{http://purl.org/np/RAhnnsMWVM8M29NixCJfVDLWzRzwwCPnUD7LI2kxT-FME#relative-neocortex-size}{relative neocortex size}* & \href{https://w3id.org/linkflows/superpattern/terms/generallyQualifierhttps://w3id.org/linkflows/superpattern/terms/neverQualifier}{never} & \href{https://w3id.org/linkflows/superpattern/terms/affects}{affects} & \href{http://purl.org/np/RAlKYv_sE8qwiSqsRdcr7KrkU1bsqlqiFmhDPtPBwpLrM#social-group-size}{social group size}* \\
        \hline
        \href{http://purl.org/np/RAoo8EvTgfkxJw5SgZXbJvRl5nQG7ygeGaHp8Zud1U4Zw}{13} & \href{http://purl.org/np/RAaOAF90U6YxAvnchfj0dRtT5HRz320Pz202aGap-VfuI#ecm-bound-cancer-cell}{ecm bound cancer cell}* & \href{http://purl.org/np/RA-jkb7qPNTSOe_EXltW_rlQWQ9x3_Y1KOzW6J_bbPz4U#glycocalyx-bulk}{glycocayx bulk}* & \href{https://w3id.org/linkflows/superpattern/terms/generallyQualifier}{generally} & \href{https://w3id.org/linkflows/superpattern/terms/increases}{increases} & \href{http://purl.org/np/RAFH8AVn-wnTcSGxvPZ1Uiy_AtOhINlynnAxxiCdcTVWU#integrin-clustering}{integrin clustering}* \\
        \hline
        \href{http://purl.org/np/RAeRSya2qIYymsBxiqOZP_oaQpHXUVXiydKvPCFM-7DDQ}{14} & \href{https://www.wikidata.org/wiki/Q5}{human (Q5)} & \href{http://purl.org/np/RAPVWYH0x-xyDa9PfBcGUFly3m1FNEO43KG9s0uH-y6yo#STX1B-mutation}{STX1B mutation}* & \href{https://w3id.org/linkflows/superpattern/terms/frequentlyQualifier}{frequently} & \href{https://w3id.org/linkflows/superpattern/terms/coOccursWith}{co-occurs with} & \href{https://www.wikidata.org/wiki/Q41571}{epilepsy (Q41571)} \\
        \hline
        \href{http://purl.org/np/RAgoIxfXPqNDY8vnK2EmBQDAFwuFIDJtfaMplTvPMq3pg}{15} $\diamond$ & \href{http://purl.org/np/RAkCjYmMU3obIrC4IpwUw84JW1ymd312yz0N0g-R9yes0#digital-humanities-research}{digital humanities research}* & \href{http://purl.org/np/RAcPa1aO8kAt8QYwjQoJq-PIzYvo0jCzYrAiOX_XOyk1w#usage-of-linked-data-scopes}{usage of Linked Data Scopes}* & \href{https://w3id.org/linkflows/superpattern/terms/canGenerallyQualifier}{can generally} & \href{https://w3id.org/linkflows/superpattern/terms/contributesTo}{contributes to} & \href{https://www.wikidata.org/wiki/Q535347}{transparency (Q535347)} \\
        \hline
    \end{tabular}
    \caption{Instantiated super-patterns accepted for publication in formalization papers in the Data Science special issue. Submissions marked with $\diamond$ are formalizations in which authors extracted a scientific claim from their own previously published article; classes minted using Nanobench are marked with *, while newly minted Wikidata classes are marked with **.}
    \label{tab:submissions-superpattern}
\end{table}

\subsection{Analysis of Nanopublications}

In Table \ref{tab:nanopubs-stats} shows the statistics about the nanopublications created during our field study\footnote{\url{https://github.com/LaraHack/formalization_papers_supplemental/tree/main/nanopubs}}. It shows a total of 15 submissions with their 15 corresponding super-pattern definitions. There are 25 updated super-patterns, indicating that some of the submissions were updated more than once. 34 new classes were minted in nanopublications as class definitions, which were subsequently used in the formalizations. 

With regard to the reviews received and the author responses, class definitions received an average number of around 3 reviews per class, while the super-pattern definitions had almost 8 review comments on average. In terms of the responses given to these reviews, the average responses to class definitions was a little over 2, while the average number of responses to the review comments for the super-pattern definitions was about 6.7.

\begin{table}[tbp]
\centering
\begin{tabular}{r|l|r|r}
icon & type & per submission & total \\
\hline
\includegraphics[width=3.5mm,trim=1mm 1mm 1mm 0,clip]{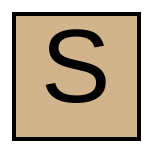} & submissions & 1.00 & 15 \\
\includegraphics[width=3.5mm,trim=1mm 1mm 1mm 0,clip]{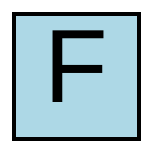} & super-pattern definitions & 1.00 & 15 \\
\includegraphics[width=3.5mm,trim=1mm 1mm 1mm 0,clip]{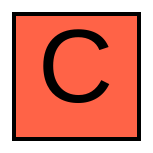} & class definitions & 2.27 & 34 \\
\includegraphics[width=3.5mm,trim=1mm 1mm 1mm 0,clip]{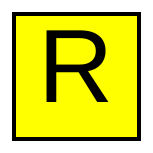} & reviews of super-patterns & 7.93 & 119 \\
\includegraphics[width=3.5mm,trim=1mm 1mm 1mm 0,clip]{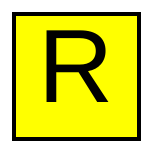} & reviews of class definitions & 3.07 & 46 \\
\includegraphics[width=3.5mm,trim=1mm 1mm 1mm 0,clip]{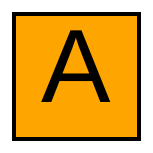} & responses to super-pattern reviews & 6.67 & 100 \\
\includegraphics[width=3.5mm,trim=1mm 1mm 1mm 0,clip]{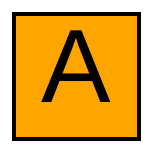} & responses to class definition reviews & 2.27 & 34 \\
\includegraphics[width=3.5mm,trim=1mm 1mm 1mm 0,clip]{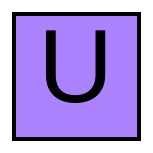} & updated super-pattern definitions & 1.67 & 25 \\
\includegraphics[width=3.5mm,trim=1mm 1mm 1mm 0,clip]{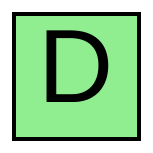} & decisions & 1.00 & 15 \\
\end{tabular}
\caption{Nanopublications created during the field study of the special issue with formalization papers at the data Science journal.}
\label{tab:nanopubs-stats}
\end{table}

\begin{figure}[tb]
\centering
\includegraphics[width=\textwidth]{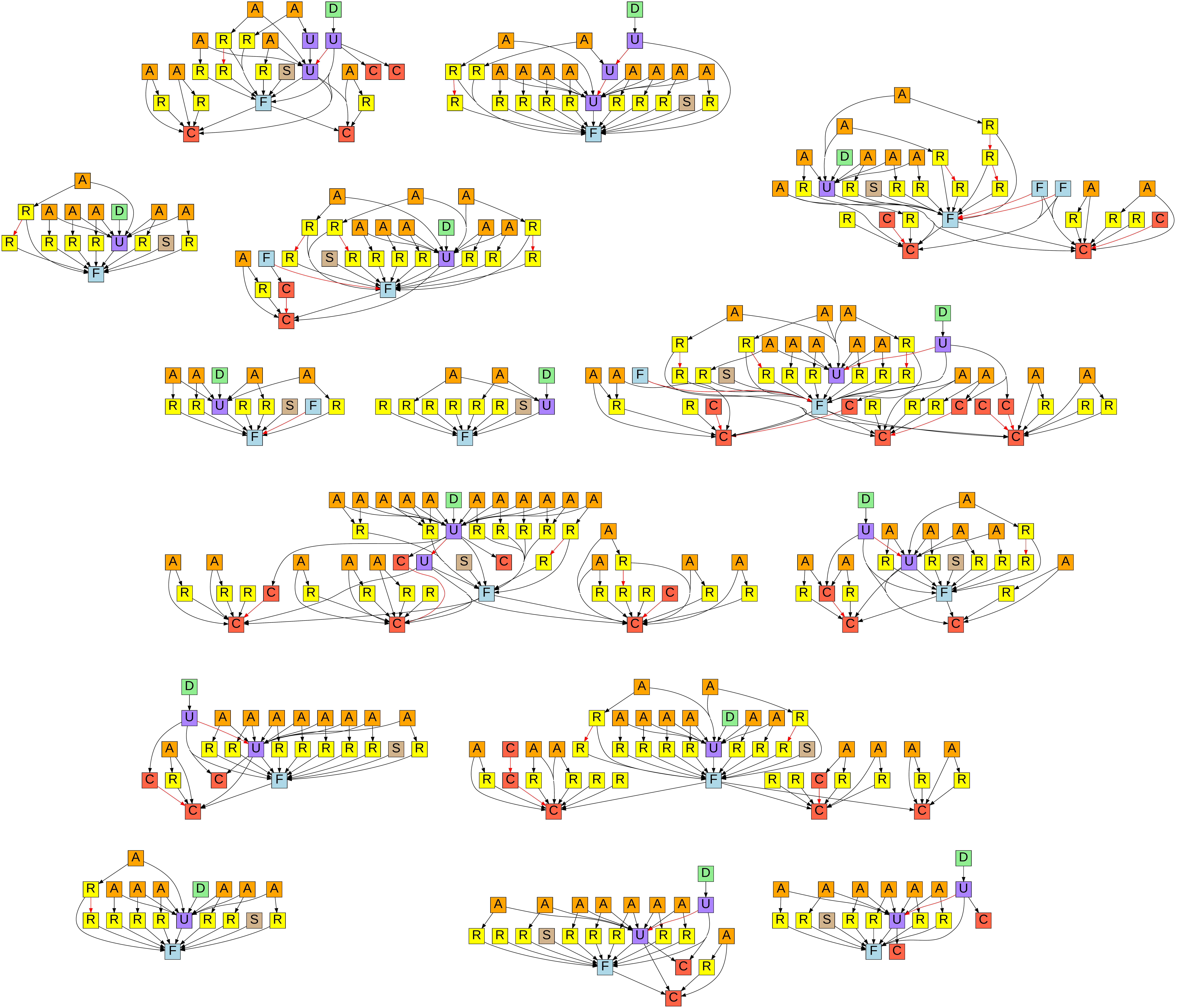}
\caption{Graphical representation of all the submissions to the Data Science special issue with formalization papers. A click-able version with links to the nanopublications can be found online: \url{https://raw.githubusercontent.com/LaraHack/fpsi_analytics/main/np-graph.svg}}
\label{fig:npgraph}
\end{figure}

In Figure \ref{fig:npgraph} we can see this graphical representation of all the special issue nanopublications, where each node represents such a nanopublication and the arrows between the nodes show how the nanopublications are linked semantically with each other. The legend for the node types indicated by color and letter code can be found in Table \ref{tab:nanopubs-stats}.

For every formalization paper, we see a first formalization (F) together with a submission nanopublication (S). Later updated versions (U) of formalizations also link back to the initial formalization. The initial submissions received review comments (R), to which authors then answered with response nanopublications (A). Additionally, some of the formalization papers used newly minted classes (C), which then also received review comments and responses. The final decision (D) points to the finally accepted updated formalization. The edges (i.e. arrows) of the graph indicate when a nanopublication is referring to another one by using its identifier in the assertion. The edges shown in red are \emph{superseding} relations, pointing from a new version of a nanopublication to its previous version. This is how nanopublications, being immutable, are dealing with representing new versions.

\subsection{User Feedback Analysis}

The 18 authors and co-authors of the formalization papers were asked to fill in the user feedback questionnaire. It was important for this questionnaire to be fully and reliably anonymous, as the authors needed to be able to give their honest opinions. This meant that we had to send reminders without knowing who already filled it in. After several rounds of reminders, we ended up getting 19 responses, meaning that at least one of the authors submitted two responses. Due to the anonymous nature of this questionnaire, it was not possible find out which responses were affected, and we have therefore to deal with such a dataset of slightly imperfect representation.

\begin{figure}[tbp]
\footnotesize
{\sffamily How difficult or easy was it for you to CONCEPTUALLY understand ...}
\vspace{-2mm}
\begin{flushright}
\scalebox{0.8}{
\begin{bchart}[max=5,min=1,step=1,width=5cm,scale=0.8]
\bcbar[label={... what a formalization paper is?}]{4.32}
\bcbar[label=... the purpose of the super-pattern?]{3.89}
\bcbar[label={... the role of the context class?}]{3.58}
\bcbar[label={... the role of the subject class?}]{3.89}
\bcbar[label={... the super-pattern qualifiers ("generally", etc.)?}]{3.68}
\bcbar[label={... the super-pattern relations ("causes", etc.)?}]{3.89}
\bcbar[label={... the role of the object class?}]{3.95}
\bcbar[label={... the overall interpretation of the super-pattern?}]{3.74}
\end{bchart}
}%
\end{flushright}
\vspace{-2mm}
{\sffamily How difficult or easy was it ...}
\vspace{-3mm}
\begin{flushright}
\scalebox{0.8}{
\begin{bchart}[max=5,min=1,step=1,width=5cm,scale=0.8]
\bcbar[label={... to find an article with a claim to formalize?}]{3.74}
\bcbar[label={... to understand what the claim exactly meant?}]{3.74}
\bcbar[label={... to decide on the context class?}]{2.89}
\bcbar[label={... to decide on the subject class?}]{2.79}
\bcbar[label={... to decide on the super-pattern qualifier?}]{3.16}
\bcbar[label={... to decide on the super-pattern relation?}]{3.11}
\bcbar[label={... to decide on the object class?}]{3.16}
\bcbar[label={... to conceptually represent the claim with the super-pattern overall?}]{2.79}
\bcxlabel{$\leftarrow$ difficult / easy $\rightarrow$}
\end{bchart}
}\vspace{-4mm}%
\end{flushright}
\caption{Questionnaire Part 1: Conceptual Aspects.}
\label{fig:questionnaire_part_1}
\end{figure}

\begin{figure}[tbp]
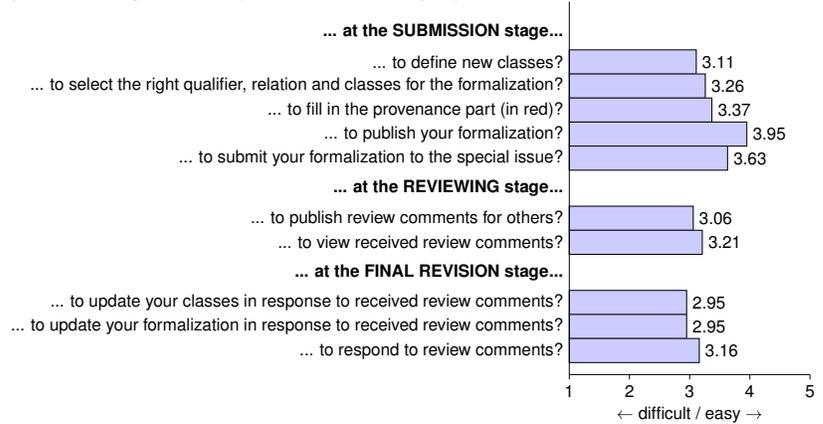

\footnotesize
{\sffamily How difficult or easy was it for you ...}
\vspace{-2mm}
\begin{flushright}
\scalebox{0.8}{
\begin{bchart}[max=5,min=1,step=1,width=5cm,scale=0.8]
\bcbar[label={... to setup Nanobench?}]{3.84}
\bcbar[label={... to use Nanobench?}]{3.32}
\bcbar[label={... to use Tapas?}]{2.76}
\end{bchart}
}%
\end{flushright}
\vspace{-2mm}
{\sffamily How difficult or easy was it for you with the given tools (Nanobench and Tapas) ...}
\vspace{-3mm}
\begin{flushright}
\scalebox{0.8}{
\begin{bchart}[max=5,min=1,step=1,width=5cm,scale=0.8]
\bigskip[label={\textbf{... at the SUBMISSION stage...}}]
\bcbar[label={... to define new classes?}]{3.11}
\bcbar[label={... to select the right qualifier, relation and classes for the formalization?}]{3.26}
\bcbar[label={... to fill in the provenance part (in red)?}]{3.37}
\bcbar[label={... to publish your formalization?}]{3.95}
\bcbar[label={... to submit your formalization to the special issue?}]{3.63}
\bigskip[label={\textbf{... at the REVIEWING stage...}}]
\bcbar[label={... to publish review comments for others?}]{3.06}
\bcbar[label={... to view received review comments?}]{3.21}
\bigskip[label={\textbf{... at the FINAL REVISION stage...}}]
\bcbar[label={... to update your classes in response to received review comments?}]{2.95}
\bcbar[label={... to update your formalization in response to received review comments?}]{2.95}
\bcbar[label={... to respond to review comments?}]{3.16}
\bcxlabel{$\leftarrow$ difficult / easy $\rightarrow$}
\end{bchart}
}\vspace{-4mm}%
\end{flushright}
\caption{Questionnaire Part 2: Technical Aspects.}
\label{fig:questionnaire_part_2}
\end{figure}

\begin{figure}[tbp]
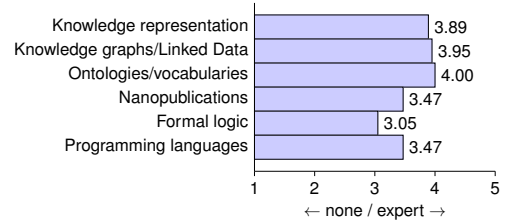

\footnotesize
{\sffamily General aspects:}
\vspace{-3mm}
\begin{flushright}
\scalebox{0.8}{
\begin{bchart}[max=5,min=1,step=1,width=5cm,scale=0.8]
\bcbar[label={Confidence as author in the quality of the formalization}]{4.00}
\bcbar[label={Importance that authors create final formalization themselves}]{3.68}
\bcbar[label={Interest in publishing such formalizations together with future articles}]{4.05}
\smallskip
\bcbar[label={Importance of ``classical view'' for website visitors}]{4.68}
\bcbar[label={Importance of ``nanopublication view'' for website visitors}]{3.32}
\bcxlabel{$\leftarrow$ not at all / very $\rightarrow$}
\end{bchart}
}%
\end{flushright}
\vspace{-2mm}
{\sffamily How would you rate your knowledge with respect to the following topics?}
\vspace{-3mm}
\begin{flushright}
\smallskip
\scalebox{0.8}{
\begin{bchart}[max=5,min=1,step=1,width=5cm,scale=0.8]
\bcbar[label={Knowledge representation}]{3.89}
\bcbar[label={Knowledge graphs/Linked Data}]{3.95}
\bcbar[label={Ontologies/vocabularies}]{4.00}
\bcbar[label={Nanopublications}]{3.47}
\bcbar[label={Formal logic}]{3.05}
\bcbar[label={Programming languages}]{3.47}
\bcxlabel{$\leftarrow$ none / expert $\rightarrow$}
\end{bchart}
}\hspace{2.1mm}%
\vspace{-4mm}%
\end{flushright}
\caption{Questionnaire Parts 3 and 4.}
\label{fig:questionnaire_parts_3_and_4}
\end{figure}

In Figure \ref{fig:questionnaire_part_1} we see the results for the first part of the questionnaire. Authors expressed that it was rather easy to understand what a formalization paper is (with a score of 4.32 out of 5). The elements of the super-pattern were found a bit harder to understand but still quite easy, with scores above 3.50. Finding an article from which to select a claim to formalize and to understand what the chosen claim really meant was also deemed easy, with scores of 3.74. The actual instantiation of the super-pattern with all its fields given the chosen claim was considered a little more difficult, with scores around 3.0, indicating medium difficulty roughly in the middle of \emph{very difficult} and \emph{very easy}. These results seem to suggest that the authors were able to understand the main formalization papers idea together with the super-pattern that comprises it, but when it came to the actual instantiation of the super-pattern (especially concerning the context and subject class), this was considered a little more difficult, but still on average far from very difficult.

In Figure \ref{fig:questionnaire_part_2}, we see the authors' responses with respect to technical difficulty. In terms of the tools used, we see that setting up and using Nanobench was considered easy enough (with a score of 3.30), while the Tapas interface seems a little harder to use (with a score of 2.76).
The different tasks in the different stages all seemed to be between medium and easy on average, with the exception of the tasks to provide responses to reviews, which scored slightly below 3.0. The response nanopublications are indeed among the most complex ones, as they refer not only to the affected review but also to the updated formalization. Overall, while these results show room for improvement, they still seem favorable given that we were building upon generic and powerful tools without specific user interface design or polishing.

Figure \ref{fig:questionnaire_parts_3_and_4} summarizes the assessment of more general aspects of formalization papers and also contains information about the authors' background. We see that authors have a high confidence in the quality of their formalization, with an average score of 4.0, and that they are interested in the future publication of such a formalization along their scientific publications, with a score of 4.05. The respondents very clearly stated that the classical view of formalization papers is important for website visitors, with a score of 4.68. Exposing also the ``naked'' nanopublications to the website visitors with a nanopublication view was found to be much less important (3.32).

The authors indicated that they have, on average, a high level of knowledge on the topics of knowledge representation, knowledge graphs, Linked Data, and ontologies/vocabularies, with scores from 3.89 to 4.00. Their background in nanopublications, formal logic, and programming languages was significantly lower, on average, but still relatively high, with scores between 3.05 and 3.47.

10 out of the 18 authors used the free text feedback of the questionnaire. 8 of these 10 respondents expressed their excitement about the field study and found the formalization paper concept and the whole publication process interesting and useful. However, half of these respondents also mentioned that the overall process proved to be a little more difficult than they expected, due to the tools used maybe being too technical. One author also pointed out that multiple formalizations can be written for the same claim by choosing the context, subject and object classes differently and expressed the worry that this would decrease the interoperatibility or utility of formalizations especially when aggregating or mining them. This is a reasonable point to make, but due to the fully formal semantics, syntactic differences are in principle not hindering this kind of interoperability. Overall, the super-pattern, the formalization paper concept, and the nanopublication-based publication workflow seem to have been well-accepted and understood by the participants, and many of them showed an enthusiastic reaction. 

\section{Discussion and Conclusion} \label{conclusion}

The publication of the special issue with formalization papers at the Data Science journal shows not only that nanopublications and the super-pattern can be used to implement the basic steps and entities of a journal workflow, but also that authors of such formalization papers can be taught to use these in order to publish in a novel journal publication workflow as the publication of the special issue demonstrates. Our results show that the super-pattern can be well understood conceptually and despite the fact that from a practical standpoint applying it seems to be more difficult, its application remains perfectly feasible. Furthermore, we saw in our field study that even if the current general-purpose tools can be considered a viable solution, these are not necessarily easy to use, but they still remain a good tool for the purpose of publishing formalization papers. Moreover, considering the formalization papers, authors seem confident with regard to the quality of their publications and seem interested in publishing such formalizations in the future.

In future work, we plan to take the next logical step by publishing novel claims in this way form the start, and not depend on claims from already-published papers. These contributions will then also have to be accompanied by statements about the methods, equipment, and all other relevant scientific concepts, and can include not just the high-level claim but more lower-level ones, possibly all the way down to the raw data. This representation would then ideally cover the entire scientific workflow, starting from a motivation, leading to the design and execution of a study, and ending in new scientific insights. Such fully formalized scientific contributions can be seen as a major step --- even a breakthrough --- for the Semantic Web and Open Science movements and will bring us closer to a world where machines can interpret scientific knowledge and help us organize and understand it in a reliable and transparent manner.

\paragraph* \noindent \textbf{Acknowledgements.}
This research was partly funded by IOS Press and the Netherlands Institute for Sound and Vision. The authors would like to thank Stephanie Delbeque, Maarten Fr\"{o}hlich and Johan Oomen for providing their insight and expertise.

\bibliography{references}

\begin{thebibliography}{}

\bibitem[Al-Moslmi et~al., 2020]{AlMoslmi2020}
Al-Moslmi, T., na, M. G.~O., Opdahl, A.~L., and Veres, C. (2020).
\newblock Named entity extraction for knowledge graphs: A literature overview.
\newblock {\em IEEE Access}, 8:32862--32881.

\bibitem[Bechhofer et~al., 2013]{Bechhofer2013RO}
Bechhofer, S., Buchan, I., {De Roure}, D., Missier, P., Ainsworth, J., Bhagat,
  J., Couch, P., Cruickshank, D., Delderfield, M., Dunlop, I., Gamble, M.,
  Michaelides, D., Owen, S., Newman, D., Sufi, S., and Goble, C. (2013).
\newblock Why linked data is not enough for scientists.
\newblock {\em Future Generation Computer Systems}, 29(2):599--611.
\newblock Special section: Recent advances in e-Science.

\bibitem[Belhajjame et~al., 2015]{Belhajjame2015RO}
Belhajjame, K., Zhao, J., Garijo, D., Gamble, M., Hettne, K., Palma, R., Mina,
  E., Corcho, O., Gómez-Pérez, J.~M., Bechhofer, S., Klyne, G., and Goble, C.
  (2015).
\newblock Using a suite of ontologies for preserving workflow-centric research
  objects.
\newblock {\em Journal of Web Semantics}, 32:16--42.

\bibitem[Benda and Engels, 2011]{Benda2011}
Benda, W.~G. and Engels, T. C.~E. (2011).
\newblock The predictive validity of peer review: A selective review of the
  judgmental forecasting qualities of peers, and implications for innovation in
  science.

\bibitem[Brack et~al., 2020]{Brack2020}
Brack, A., D'Souza, J., Hoppe, A., Auer, S., and Ewerth, R. (2020).
\newblock Domain-independent extraction of scientific concepts from research
  articles.
\newblock In {\em Advances in Information Retrieval}, pages 251--266, Cham.
  Springer International Publishing.

\bibitem[{Bucur} et~al., 2019]{Bucur2019ReviewModel}
{Bucur}, C.-I., {Kuhn}, T., and {Ceolin}, D. (2019).
\newblock Peer reviewing revisited: Assessing research with interlinked
  semantic comments.
\newblock In {\em In K-CAP 2019: Proceedings of the 10th International
  Conference on Knowledge Capture}, pages 179--187.

\bibitem[Bucur et~al., 2020]{Bucur2020}
Bucur, C.-I., Kuhn, T., Ceolin, D., and van Ossenbruggen, J. (2020).
\newblock A unified nanopublication model for effective and user-friendly
  access to the elements of scientific publishing.
\newblock {\em EKAW2020}.

\bibitem[Bucur et~al., 2021]{Bucur2021SP}
Bucur, C.-I., Kuhn, T., Ceolin, D., and van Ossenbruggen, J. (2021).
\newblock Expressing high-level scientific claims with formal semantics.
\newblock In {\em Proceedings of the 11th on Knowledge Capture Conference},
  K-CAP '21, page 233–240, New York, NY, USA. Association for Computing
  Machinery.

\bibitem[Chi et~al., 2018]{Chi2018KnowledgeGI}
Chi, Y., Qin, Y., Song, R., and Xu, H. (2018).
\newblock Knowledge graph in smart education: A case study of entrepreneurship
  scientific publication management.
\newblock {\em Sustainability}, 10:995.

\bibitem[Chibucos et~al., 2014]{Chibucos2014}
Chibucos, M.~C., Mungall, C.~J., Balakrishnan, R., Christie, K.~R., Huntley,
  R.~P., White, O., Blake, J.~A., Lewis, S.~E., and Giglio, M. (2014).
\newblock Standardized description of scientific evidence using the evidence
  ontology (eco).
\newblock {\em Database : the journal of biological databases and curation}.

\bibitem[Coulet et~al., 2011]{Coulet2011}
Coulet, A., Garten, Y., Dumontier, M., Altman, R.~B., Musen, M.~A., and Shah,
  N.~H. (2011).
\newblock Integration and publication of heterogeneous text-mined relationships
  on the semantic web.
\newblock {\em J Biomed Semant}, 2.

\bibitem[Etzioni et~al., 2005]{Etzioni2005}
Etzioni, O., Cafarella, M., Downey, D., Popescu, A.-M., Shaked, T., Soderland,
  S., Weld, D.~S., and Yates, A. (2005).
\newblock Unsupervised named-entity extraction from the web: An experimental
  study.
\newblock {\em Artificial Intelligence}, 165(1):91--134.

\bibitem[F{\"a}rber and Lamprecht, 2021]{Frber2021TheDS}
F{\"a}rber, M. and Lamprecht, D. (2021).
\newblock The data set knowledge graph: Creating a linked open data source for
  data sets.
\newblock {\em Quantitative Science Studies}.

\bibitem[Felix and Barrand, 2002]{Felix2002}
Felix, R.~A. and Barrand, M.~A. (2002).
\newblock P-glycoprotein expression in rat brain endothelial cells: evidence
  for regulation by transient oxidative stress.
\newblock {\em Journal of Neurochemistry}, 80:64--72.

\bibitem[Garcia-Castro et~al., 2013]{GarciaCastro2013}
Garcia-Castro, L., Berlanga, R., Rebholz-Schuhmann, D., and Garcia, A. (2013).
\newblock Connections across scientific publications based on semantic
  annotations.
\newblock {\em SEPublica, 10th Extended Semantic Web Conference}.

\bibitem[Garijo and Poveda-Villal\'on, 2020]{Garijo2020BestPF}
Garijo, D. and Poveda-Villal\'on, M. (2020).
\newblock Best practices for implementing fair vocabularies and ontologies on
  the web.
\newblock In {\em Applications and Practices in Ontology Design, Extraction,
  and Reasoning}.

\bibitem[Groth et~al., 2010]{Groth2010}
Groth, P., Gibson, A., and Velterop, J. (2010).
\newblock The anatomy of a nanopublication.
\newblock {\em Inf. Serv. Use}, 30:51--56.

\bibitem[Halpin et~al., 2010]{Halpin2010}
Halpin, H., Hayes, P.~J., McCusker, J.~P., McGuinness, D.~L., and Thompson,
  H.~S. (2010).
\newblock When owl:sameas isn't the same: An analysis of identity in linked
  data.
\newblock In {\em The Semantic Web -- ISWC 2010}, pages 305--320, Berlin,
  Heidelberg. Springer Berlin Heidelberg.

\bibitem[Hannestad et~al., 2021]{Hannestad2021KnowledgeBW}
Hannestad, L.~M., Danc{\'i}k, V., Godden, M., Suen, I.~W., Huellas-Bruskiewicz,
  K.~C., Good, B.~M., Mungall, C.~J., and Bruskiewich, R.~M. (2021).
\newblock Knowledge beacons: Web services for data harvesting of distributed
  biomedical knowledge.
\newblock {\em PLoS ONE}, 16.

\bibitem[Hitzler and van Harmelen, 2010]{Hitzler2010}
Hitzler, P. and van Harmelen, F. (2010).
\newblock A reasonable semantic web.
\newblock {\em Semantic Web}, 1:39--44.

\bibitem[Hyv{\"o}nen, 2012]{Hyvnen2012PublishingAU}
Hyv{\"o}nen, E. (2012).
\newblock Publishing and using cultural heritage linked data on the semantic
  web.
\newblock In {\em Synthesis Lectures on the Semantic Web}.

\bibitem[Hyv{\"o}nen, 2020]{Hyvnen2020UsingTS}
Hyv{\"o}nen, E. (2020).
\newblock Using the semantic web in digital humanities: Shift from data
  publishing to data-analysis and serendipitous knowledge discovery.
\newblock {\em Semantic Web}, 11:187--193.

\bibitem[Jacob et~al., 2017]{Jacob2017DataworldAP}
Jacob, B., Griffith, D., and Le, T.~Q. (2017).
\newblock Data.world: A platform for global-scale semantic publishing.
\newblock In {\em SEMWEB}.

\bibitem[Jain et~al., 2010]{Jain2010}
Jain, P., Hitzler, P., Yeh, P.~Z., Verma, K., and Sheth, A.~P. (2010).
\newblock Linked data is merely more data.
\newblock In {\em Linked Data Meets Artificial Intelligence}, page 82–86.
  AAAI.

\bibitem[Jaradeh et~al., 2019]{Jaradeh2019}
Jaradeh, M.~Y., Oelen, A., Farfar, K.~E., Prinz, M., D'Souza, J., Kismih\'ok,
  G., Stocker, M., and Auer, S. (2019).
\newblock Open research knowledge graph: Next generation infrastructure for
  semantic scholarly knowledge.
\newblock In {\em Proceedings of the 10th International Conference on Knowledge
  Capture}, K-CAP'19, page 243–246, New York, NY, USA. Association for
  Computing Machinery.

\bibitem[Kotturi et~al., 2017]{Kotturi2017}
Kotturi, Y., Du, A., Klemmer, S., and Kulkarni, C. (2017).
\newblock Long-term peer reviewing effort is anti-reciprocal.
\newblock In {\em Proceedings of the Fourth (2017) ACM Conference on Learning @
  Scale}, L@S '17, page 279–282, New York, NY, USA. Association for Computing
  Machinery.

\bibitem[{Kuhn} et~al., 2013]{kuhn2013broadening}
{Kuhn}, T., {Barbano}, P.~E., {Nagy}, M.~L., and {Krauthammer}, M. (2013).
\newblock Broadening the scope of nanopublications.
\newblock In {\em Extended Semantic Web Conference}, pages 487--5017.

\bibitem[Kuhn and Dumontier, 2015]{kuhn2015making}
Kuhn, T. and Dumontier, M. (2015).
\newblock Making digital artifacts on the web verifiable and reliable.
\newblock {\em IEEE Transactions on Knowledge and Data Engineering},
  27(9):2390--2400.

\bibitem[Kuhn and Dumontier, 2017]{Kuhn2017GenuineSP}
Kuhn, T. and Dumontier, M. (2017).
\newblock Genuine semantic publishing.
\newblock {\em Data Sci.}, 1:139--154.

\bibitem[Kuhn et~al., 2021]{kuhn2021semantic}
Kuhn, T., Taelman, R., Emonet, V., Antonatos, H., Soiland-Reyes, S., and
  Dumontier, M. (2021).
\newblock Semantic micro-contributions with decentralized nanopublication
  services.
\newblock {\em PeerJ Computer Science}, 7:e387.

\bibitem[Lee et~al., 2012]{Lee2012}
Lee, C.~J., Sugimoto, C.~R., Zhang, G., and Cronin, B. (2012).
\newblock Bias in peer review.
\newblock {\em Journal of the American Society for Information Science and
  Technology}.

\bibitem[Linkov et~al., 2006]{Linkov2006}
Linkov, F., Lovalekar, M., and LaPorte, R. (2006).
\newblock Scientific journals are `'faith based": is there science behind peer
  review?
\newblock {\em Journal of the Royal Society of Medicine}, 99:596--598.

\bibitem[Lisena et~al., 2019]{lisena2019easy}
Lisena, P., Mero{\~n}o-Pe{\~n}uela, A., Kuhn, T., and Troncy, R. (2019).
\newblock Easy web api development with sparql transformer.
\newblock In {\em International semantic web conference}, pages 454--470.
  Springer.

\bibitem[Madan et~al., 2019]{Madan2019}
Madan, S., Szostak, J., Elayavilli, R.~K., Tsai, R. T.-H., Ali, M., Qian, L.,
  Rastegar-Mojarad, M., Hoeng, J., and Fluck, J. (2019).
\newblock The extraction of complex relationships and their conversion to
  biological expression language (bel) overview of the biocreative vi (2017)
  bel track.
\newblock {\em Database : the journal of biological databases and curation}.

\bibitem[McGregor, 2008]{McGregor2008Facets}
McGregor, B. (2008).
\newblock Facets and hierarchies in scientific search.
\newblock {\em The Journal of Electronic Publishing}, 11.

\bibitem[McNutt et~al., 2018]{McNutt2018TransparencyIA}
McNutt, M.~K., Bradford, M., Drazen, J.~M., Hanson, B., Howard, B., Jamieson,
  K.~H., Kiermer, V., Marcus, E., Pope, B.~K., Schekman, R., Swaminathan, S.,
  Stang, P., and Verma, I.~M. (2018).
\newblock Transparency in authors’ contributions and responsibilities to
  promote integrity in scientific publication.
\newblock {\em Proceedings of the National Academy of Sciences of the United
  States of America}, 115:2557 -- 2560.

\bibitem[Mero{\~n}o-Pe{\~n}uela and Hoekstra, 2016]{merono2016grlc}
Mero{\~n}o-Pe{\~n}uela, A. and Hoekstra, R. (2016).
\newblock grlc makes github taste like linked data apis.
\newblock In {\em European Semantic Web Conference}, pages 342--353. Springer.

\bibitem[Papakonstantinou et~al., 2018]{Papakonstantinou2018AssessingLD}
Papakonstantinou, V., Fundulaki, I., and Flouris, G. (2018).
\newblock Assessing linked data versioning systems: The semantic publishing
  versioning benchmark.
\newblock In {\em SSWS@ISWC}.

\bibitem[Perez-Arriaga, 2018]{PerezArriaga2018AutomatedDO}
Perez-Arriaga, M. (2018).
\newblock Automated development of semantic data models using scientific
  publications.

\bibitem[Peroni, 2014]{Peroni2014TheSP}
Peroni, S. (2014).
\newblock The semantic publishing and referencing ontologies.

\bibitem[Peroni, 2017]{Peroni2017AutomatingSP}
Peroni, S. (2017).
\newblock Automating semantic publishing.
\newblock {\em Data Sci.}, 1:155--173.

\bibitem[Peroni et~al., 2016]{Peroni2016ResearchAI}
Peroni, S., Osborne, F., Iorio, A.~D., Nuzzolese, A.~G., Poggi, F., Vitali, F.,
  and Motta, E. (2016).
\newblock Research articles in simplified html: a web-first format for
  html-based scholarly articles.
\newblock {\em PeerJ Prepr.}, 4:e2513.

\bibitem[Peroni et~al., 2013]{Peroni2013Lenses}
Peroni, S., Tomasi, F., Vitali, F., and Zingoni, J. (2013).
\newblock Semantic lenses as exploration method for scholarly articles.
\newblock In T., C., N., F., and A., P., editors, {\em Italian Research
  Conference on Digital Libraries: Bridging Between Cultural Heritage
  Institutions, Communications in Computer and Information Science}, volume 385
  of {\em IRCDL'13}, Berlin, Heidelberg, Germany. Springer.

\bibitem[Rahardja et~al., 2019]{Rahardja2019ScientificPM}
Rahardja, U., Lutfiani, N., and Juniar, H.~L. (2019).
\newblock Scientific publication management transformation in disruption era.
\newblock {\em Aptisi Transactions on Management (ATM)}.

\bibitem[Sateli and Witte, 2016]{Sateli2016FromPT}
Sateli, B. and Witte, R. (2016).
\newblock From papers to triples: An open source workflow for semantic
  publishing experiments.

\bibitem[Senderov and Penev, 2016]{Senderov2016TheOB}
Senderov, V. and Penev, L. (2016).
\newblock The open biodiversity knowledge management system in scholarly
  publishing.
\newblock {\em Research Ideas and Outcomes}, 2.

\bibitem[Sernadela et~al., 2015]{Sernadela2015}
Sernadela, P., Lopes, P., Campos, D., Matos, S., and Oliveira, J.~L. (2015).
\newblock A semantic layer for unifying and exploring biomedical document
  curation results.
\newblock {\em WBBIO'2015}.

\bibitem[Shotton, 2009]{Shotton2009SemanticPT}
Shotton, D.~M. (2009).
\newblock Semantic publishing: the coming revolution in scientific journal
  publishing.
\newblock {\em Learned Publishing}, 22.

\bibitem[Shotton et~al., 2009]{Shotton2009AdventuresIS}
Shotton, D.~M., Portwin, K., Klyne, G., and Miles, A.~J. (2009).
\newblock Adventures in semantic publishing: Exemplar semantic enhancements of
  a research article.
\newblock {\em PLoS Computational Biology}, 5.

\bibitem[Slater, 2014]{Slater2014}
Slater, T. (2014).
\newblock Recent advances in modeling languages for pathway maps and computable
  biological networks.
\newblock {\em Drug Discovery Today}, 19.

\bibitem[Slater and Song, 2012]{Slater2012}
Slater, T. and Song, D.~H. (2012).
\newblock Saved by the bel: ringing in a common language for the life sciences.
\newblock {\em Fall}.

\bibitem[Smith, 1988]{Smith1988}
Smith, R. (1988).
\newblock Problems with peer review and alternatives.
\newblock {\em British Medical Journal (Clinical Research Edition)},
  296(6624):774--777.

\bibitem[Smith, 2010]{Smith2010}
Smith, R. (2010).
\newblock Classical peer review: an empty gun.
\newblock {\em Breast cancer research}, 12.

\bibitem[Tenorio-Forn{\'e}s et~al., 2019]{TenorioForns2019TowardsAD}
Tenorio-Forn{\'e}s, A., Jacynycz, V., Llop-Vila, D., S{\'a}nchez-Ruiz-Granados,
  A.~A., and Hassan, S. (2019).
\newblock Towards a decentralized process for scientific publication and peer
  review using blockchain and ipfs.
\newblock In {\em HICSS}.

\bibitem[Tiddi et~al., 2020]{Tiddi2020FosteringSM}
Tiddi, I., Balliet, D., and ten Teije, A. (2020).
\newblock Fostering scientific meta-analyses with knowledge graphs: A
  case-study.
\newblock {\em The Semantic Web}, 12123:287 -- 303.

\bibitem[Wilkinson et~al., 2016]{Wilkinson2016TheFG}
Wilkinson, M.~D., Dumontier, M., Aalbersberg, I.~J., Appleton, G., Axton, M.,
  Baak, A., Blomberg, N., Boiten, J.-W., da~Silva~Santos, L. O.~B., Bourne,
  P.~E., Bouwman, J., Brookes, A.~J., Clark, T., Crosas, M., Dillo, I., Dumon,
  O., Edmunds, S.~C., Evelo, C. T.~A., Finkers, R., Gonz{\'a}lez-Beltr{\'a}n,
  A.~N., Gray, A. J.~G., Groth, P., Goble, C.~A., Grethe, J.~S., Heringa, J.,
  ‘t Hoen, P. A.~C., Hooft, R. W.~W., Kuhn, T., Kok, R.~G., Kok, J.~N.,
  Lusher, S.~J., Martone, M.~E., Mons, A., Packer, A.~L., Persson, B.,
  Rocca-Serra, P., Roos, M., van Schaik, R.~C., Sansone, S.-A., Schultes,
  E.~A., Sengstag, T., Slater, T., Strawn, G.~O., Swertz, M.~A., Thompson, M.,
  van~der Lei, J., van Mulligen, E.~M., Velterop, J., Waagmeester, A.,
  Wittenburg, P., Wolstencroft, K., Zhao, J., and Mons, B. (2016).
\newblock The fair guiding principles for scientific data management and
  stewardship.
\newblock {\em Scientific Data}, 3.

\bibitem[Wong, 2019]{Wong2019ACO}
Wong, C.-Y. (2019).
\newblock A century of scientific publication: towards a theorization of growth
  behavior and research-orientation.
\newblock {\em Scientometrics}, 119:357--377.

\bibitem[Xu et~al., 2015]{Xu2015}
Xu, Y., Mou, L., Li, G., Chen, Y., Peng, H., and Jin, Z. (2015).
\newblock Classifying relations via long short term memory networks along
  shortest dependency paths.
\newblock In {\em Proceedings of the 2015 Conference on Empirical Methods in
  Natural Language Processing}, pages 1785--1794, Lisbon, Portugal. Association
  for Computational Linguistics.

\bibitem[Yadav and Bethard, 2018]{Yadav2018}
Yadav, V. and Bethard, S. (2018).
\newblock A survey on recent advances in named entity recognition from deep
  learning models.
\newblock In {\em COLING}.

\bibitem[Zeng et~al., 2014]{Zeng2014}
Zeng, D., Liu, K., Lai, S., Zhou, G., and Zhao, J. (2014).
\newblock Relation classification via convolutional deep neural network.
\newblock In {\em COLING}.

\end{thebibliography}

\end{document}